\documentclass[twocolumn]{revtex4}
\usepackage{graphicx}
\usepackage[]{epsfig}

\newcommand{\beq}{\begin{equation}}
\newcommand{\eeq}{\end{equation}}
\newcommand{\beqd}{\begin{displaymath}}
\newcommand{\eeqd}{\end{displaymath}}
\newcommand{\beqa}{\begin{eqnarray}}
\newcommand{\eeqa}{\end{eqnarray}}
\newcommand{\non}{\nonumber}

\newcommand{\comment}[1]{}

\begin{document}

\title{Phase diagram and large deviations in the free-energy of mean-field spin-glasses}


\author{Giorgio Parisi$^{1,2}$ and Tommaso Rizzo$^{3}$}

\affiliation{$^{1}$Dipartimento di Fisica, Universit\`a di Roma ``La Sapienza'', 
P.le Aldo Moro 2, 00185 Roma,  Italy
\\
$^{2}$Statistical Mechanics and Complexity Center (SMC) - INFM - CNR, Italy
\\
$^{3}$ ``E. Fermi'' Center, Via Panisperna 89 A, Compendio Viminale, 00184, Roma, Italy}

\begin{abstract}
    We consider the probability distribution of large deviations in the spin-glass free energy 
     for the Sherrington-Kirkpatrick mean field model, {\it i.e.}  the exponentially small
    probability of finding a system with intensive free energy smaller than the most likely one.  This result
    is obtained by computing $\Phi(n,T)=T \overline{Z^n}/ n$, {\it i.e.} the average value of the partition function to the power $n$  as a function of $n$. 
 We study in full details the phase diagram  of $\Phi(n,T)$  in the $(n,T)$ plane computing in particular the stability of the replica-symmetric solution.
 At low temperatures we compute $\Phi(n,T)$ in series of $n$ and $\tau=T_c-T$ at high orders using the standard hierarchical ansatz and confirm earlier findings on the $O(n^5)$ scaling. We prove that the $O(n^5)$ scaling is valid at all orders and obtain an exact expression for the coefficient in term of the function $q(x)$. Resumming the series we obtain the large deviations probability at all temperatures. 
At zero temperature the analytical prediction displays a remarkable quantitative agreement with the numerical data.
A similar computation for the simpler spherical model is also performed and the connection between large and small deviations is discussed.
\end{abstract}

\maketitle

\section{Introduction}

The theory of disordered systems is mainly concerned with predictions regarding the most likely behavior, but it is also interesting to develop techniques to compute the probability distribution of
rare events, i.e. the probability  of finding systems that have properties different
from the typical ones. There are various motivations:
\begin{itemize}
    \item We may have a  special interest in those systems with a behavior different from the most likely one; for example in constraint optimization problems, in the region where it is impossible to satisfy all the constraints in the most likely system, there is a great interest in computing the properties of those rare systems where we can find a configuration that satisfies all the constraints \cite{AMZ}.
   \item The properties of large fluctuations may be related to other more interesting properties of the 
    system. For example given an intensive quantity $A_J$ that depends on  the system $J$ of size $N$, in the large deviation region for large  $N$ we usually have that $P_N(A)\approx \exp (-N L(A))$. It is quite common that there are relations among the behavior of $P_N(A)$ in the region where the probability remains finite when $N$ goes to infinity and the behavior of $L(A)$ near the point where  $L(A)=0$. In other cases \cite{AM0} the techniques used to compute large deviations are the same used to compute other important quantities like (in finite dimensional spin glasses) the typical difference of the energy   with periodic and antiperiodic boundary conditions. Besides sample to sample fluctuations have been recently shown to be related to chaos in spin glasses \cite{ACHAOS}.
        \item We notice also that the comparison between analytic predictions in  the large deviations region and numerical or experimental data could be important as a clear-cut test of the theoretical approach used to compute the most likely
    properties.
 
\end{itemize}

Unfortunately even in the simplest non-trivial case, i.e. the Sherrington-Kirkpatrick (SK) infinite range
model for spin glasses, there is no consensus on the results of such a computation. Everybody agrees
that as a first step we need to compute in the large $N$ limit the thermodynamic function
\beq
\Phi(n,\beta)=-{1\over \beta n N}\ln \overline{Z_{J}(\beta)^n}\ ,
\label{defphi}
\eeq
where different systems (or samples) are labeled by $J$, $Z_{J}(\beta)$ is the partition function and the
bar denotes the average over different disordered samples.  It is well known that the probability of large
deviations is related to the function $\Phi(n,\beta)$. Indeed
\beq
\exp( -\beta n N \Phi(n,\beta))=\overline{Z_{J}(\beta)^n}= \overline{\exp(-nN \beta f_{J}(\beta))}\ ,
\eeq
where $ f_{J}$ is the system-dependent free energy {\it per spin}. The region of positive $n$ corresponds to fluctuations where the free energy is smaller than the typical one and the region on negative $n$ corresponds to fluctuations where the free energy is larger than the typical one.

There is a disagreement in the literature on the strategy we should follow to 
compute  $\Phi(n,\beta)$. In the $n \rightarrow 0$ limit the computation can be done using the broken replica symmetry ansatz (that is
known to give the exact result), where it coincides with the most likely free energy $\Phi(0,\beta)=f_{typ}$ or equivalently with the average equilibrium free energy $\overline{f_{eq}}=f_{typ}$. 

For $n>0$ Kondor \cite{Kon1} in 1983 presented a first computation of $\Phi(n,\beta)$ in the region near
$T_{c}$ using the most natural ansatz for replica symmetry breaking (RSB) obtaining in the region of positive $n$
\beq
\Phi(n,\beta)= f_{typ}+c_5n^5 + O(n^6)\ .\label{KONDOR}
\eeq
The result of Kondor was surprising: in the general case  all powers of $n$ are present in the Taylor expansion of $\Phi(n,\beta)$ and for most of the systems we have $\Phi(n,\beta)= f_{typ}+A_1n+ O(n^2)$, that is the typical situations for a Gaussian distribution of the free energy. The absence of the powers from $n^1$ to $n^4$ is due to cancellations and it was not clear if they were present only near the critical temperature. This form of the large deviation function implies that the probability distribution for $f$ near (and smaller than) $f_{typ}$ is of the form
\beq
P_N(f)\propto \exp(-Na_{6/5}(f_{typ}-f)^{6/5})\ ,
\label{LP}
\eeq
where $a_{6/5}=5 \beta 6^{-6/5} |c_5|^{-1/5} $.
The above relationship is valid for a small negative value of the free energy difference $\Delta f=f-f_{typ}$ that remains finite in the thermodynamic limit.
However it has not been possible to test directly Kondor prediction because presently all numerical data concern the
fluctuations of the ground state energy, i.e. the system is at zero temperature. Indeed at zero temperature the free energy coincides with the internal energy and the numerical data are cleaner due to the absence of thermal noise. 
Instead many efforts have been concentrated on the
scaling of the small deviations of the free energy.  Indeed based on Kondor's result and a matching argument (see discussion below) it was suggested in \cite{CPSV}  
that the small deviations from its mean of the free energy per spin scale as $N^{-5/6}$. 
This prediction has
been put to test in a series of numerical works \cite{B1,BKM,CMPP,PALA,B2,KKLJH,PAL} and although all 
estimates are smaller than $5/6$ nobody has claimed that this value is definitively ruled out.   More recent results strongly indicate that the fluctuations of the internal energy  per spin at finite temperature scale as  $N^{-5/6}$, thus confirming the exponent obtained from \cite{ABMM}.
However it was
difficult to test the theory in absence of a quantitative prediction (the only prediction being on the
exponent, a quantity that it is rather difficult to measure in a reliable way).
Furthermore, the potential $\Phi(n,beta)$ is naturally related to large deviations while the matching argument connection with the small deviation exponent is not rigorous, see discussion below.

More recently a
different RSB ansatz was proposed by Aspelmeier and Moore
\cite{AM,DDF}, who found $\Phi(n)=f_{typ}$; in their approach the probability of large deviations goes to zero
faster than $\exp(-L(f) N)$ and the matching argument cannot be used to infer the small deviations exponent.
 Indeed there is a general agreement   that for negative $n$ $\Phi(n)=f_{typ}$ and $P_N(f)$ goes to zero faster than $\exp( -CN)$ as soon $f>f_{typ}$. Recent results \cite{PRprep} show that in that region we have $P_N(f)\propto\exp(-N^2 L_2(f))$ were the function $L_2(f)$ can be computed through the replica method.
This $O(N^2)$ scaling of the logarithm of the large deviations probability in the positive $\Delta f$ region is also observed in the spherical model where it has been recently derived using random matrix theory \cite{DM}.

We have  concentrated on large deviations in the region  $f<f_{typ}$, that corresponds to positive $n$. We have followed Kondor's approach and
extended his computation to all temperatures, including $T=0$; in this way we have obtained an absolute
prediction for the large deviations distribution. Comparing our analytic results with the
numerical simulations performed at zero temperature we found a remarkable agreement.
We have worked in perturbation theory assuming small $\tau=T-T_c$ and small $n$ and used appropriate resummation techniques to extend the computation down to zero temperature.
We have also verified analytically that the $O(n^5)$ scaling holds at all orders in perturbation theory and obtained an exact relationship between the corresponding coefficient and the derivative at $x=0$ of the standard $q(x)$ function.

We also
found  that the
alternative approach \cite{AM,DDF} that predicts $\Phi(n)=f_{typ}$ for both negative and positive values of $n$ cannot be valid for large positive $n$ and there are no compelling reasons for which it 
should be valid at fixed positive $n$ when $N$ goes to infinity. This is in agreement with the results coming from an  exact analysis: for positive values of $n$ Talagrand \cite{TALA} was
able to show rigorously that Kondor's approach gives  the correct results. 

The paper is organized as follows. In section \ref{SC} we introduce the functionals and the saddle point equations.
In section \ref{RS} we discuss the Replica-Symmetric (RS) solution and its stability, we compute the DeAlmeida-Thouless line in the $(n,T)$ plane and we discuss the behaviour of the sample complexity above the critical temperature.
In section \ref{RSB} we discuss the sample complexity in the low temperature phase and compare it with the numerical data. 
In section \ref{SPHERICAL} we present a similar treatment of the spherical model, which being RS is considerably simpler.
In section \ref{small} we discuss the connection between large and small deviations. 
In the last section we give our conclusions.
In appendix \ref{appa} we report the power series of $\Phi(n,T)$ up to the 18th order.
In appendix \ref{xplat} we present an analytical argument to prove that the $O(n^5)$ scaling is valid at all orders in perturbation theory and an exact relationship between the coefficient and the derivative $dq/dx$ in $x=0$.
A brief report on these results has been given in \cite{PR1}.

\section{Sample Complexity}
\label{SC}
We define the large deviation function for the free energy, $L(f)$, (that we will call in the following  the sample complexity because it is related to the number of samples with free energy equal to $f$) as the logarithm divided by $N$ of the probability density of samples with free energy per spin $f$  in the thermodynamic limit:
\beq
L (f) =\lim_{N\to\infty}{\log(P_{N}(f)) \over N }\ .
\eeq

 For large $N$ the majority of the samples has free energy per spin equal to $f_{typ}$, and all other values have exponentially small probability. Consistently $L(f)$ is less or equal than zero, the equality holding $f=f_{typ}$, i.e. $L (f_{typ})=0$. For some values of $f$ it is possible that $L (f)=-\infty$,
meaning that the probability of large deviations goes to zero faster than exponentially with $N$.
 In the thermodynamic limit the  function $\Phi(n,\beta)$ defined in eq. (\ref{defphi}) yields the Legendre transform of $L(f)$ \cite{CPSV}, indeed we have:
\beq
-\beta n \Phi(n)=- \beta n f+ L(f)
\label{prima}
\eeq
where $f$ is determined by the condition:
\beq
\beta n={\partial L \over \partial f} 
\label{prima2}
\eeq
and equivalently we have:
\beq
 L(f)=\beta n f-\beta n \Phi(n)
\label{prima3}
\eeq
where $\beta n$ is determined by the condition:
\beq
f={\partial n \Phi \over \partial n} .
\label{prima4}
\eeq
Note that at any finite $N$, $n \Phi(n)$ is also the generating function of the cumulants of the distribution of the free energy.  

\begin{figure}[htb]
\begin{center}
\epsfig{file=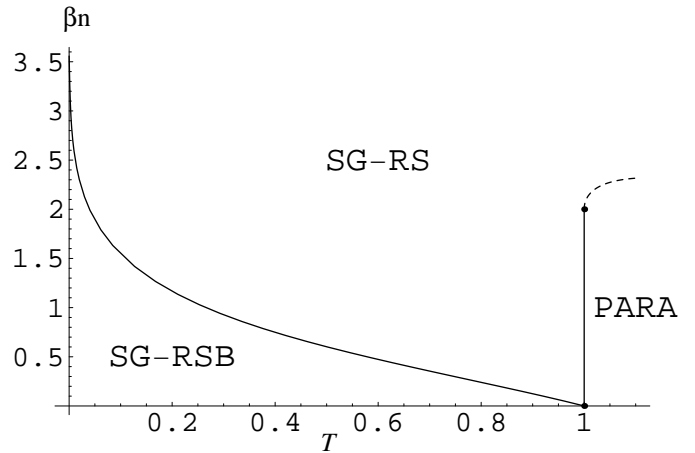,width=9cm}
\caption{Phase Diagram of $\Phi(n)$ in the $(T,\beta n)$ plane. In the paramagnetic phase the solution is RS with $q=0$ and $\Phi(n,\beta)=-\beta /4-\ln 2/\beta$. The dashed line $\beta n_c(T)$ marks a first order transition to a RS spin-glass phase where $q$ jumps from zero to a finite value $q_c(T)$. The line $n_c(T)$ ends at the point $(T=1,n=2)$ where $q_c=0$. The vertical line from the point $(1,0)$ to $(1,2)$ marks a phase transition from a paramagnetic to a RS spin-glass phase with the parameter $q$ changing continuously. The dAT line marks the region of stability of the RS spin-glass phase, below the line the phase is RSB.  The value of $\beta n_{dAT}$ diverges in the zero-temperature limit as $\beta n\simeq \sqrt{-2 \ln [3 (\pi/2)^{1/2} T]}$, as a consequence the function $L(\Delta e)$ at zero temperature is described by the RSB solution at any value of $\Delta e$.}
\label{dAT}
\end{center}\end{figure}

In the Sherrington-Kirkpatrick model at low temperatures the replica symmetry is spontaneously broken for the generic system, i.e. in the $n \rightarrow 0$ limit. One knows that at high positive values of $n$, replica symmetry is not broken \cite{S}. Therefore for positive $n$ one must distinguish two regions in the $T-n$ plane separated by the so called
de Almeida Thouless (dAT) line, see fig. (\ref{dAT}). In the region above the dAT line, the phase is replica-symmetric, while replica symmetry is broken below.

\section{The Replica-Symmetric Phase}
\label{RS}
In the Replica-Symmetric (RS) region the order parameter is the overlap $q$.  The corresponding value of the
potential $\Phi(n,q)$ is given by
\beqa
    \Phi(n,\beta)[q] & = &  -{\beta \over 4}\left(1- 2 q+(1-n) q^2\right)
+
\non
\\
& - & {1 \over \beta n}\ln \int_{-\infty}^{+\infty}{dy \over \sqrt{2 \pi q}}e^{-{y^2\over 2
 q}}(2\cosh \beta y)^n \ .
\non
\eeqa
The overlap $q$ can be computed by solving the  equation $\partial \Phi(n,q)/ \partial q =0$ that yields:
\beq
q={ 
\int e^{-{y^2\over 2 q}}(\cosh \beta y)^n \tanh^2 \beta y \, dy
\over
\int e^{-{y^2\over 2 q}}(\cosh \beta y)^n \, dy
}
\label{qRS}
\eeq

\subsection{The High-Temperature Region}
The search for solutions with $q \neq 0$ of equation (\ref{qRS}) at given $T$ and $n$ was done integrating numerically the r.h.s. for different values of $q$ and checking if the corresponding curve $y=y(q)$ crosses the line $y=q$. 

In the high temperature region $T>1$ the solution in the $n \rightarrow 0$ limit is replica symmetric with $q=0$. For small $n$ there is no other solution, therefore $\Phi(n)= f_{RS} \equiv -\beta/4-\ln2 / \beta$.
At $n=n^{*}(T)$ a new solution appears, but it has a value of $\Phi_{RS,q\neq 0}(n)$ larger than $f_{RS}$ and must be discarded ($\Phi(n)$ must be  a continuous function of $n$).
As soon as $n>n^{*}(T)$ the solution bifurcates into two solutions $q_>(n)>q_<(n)$ with the largest solution having a smaller value of $\Phi(n)$ than the other. 
Increasing $n$ we cross the line $n=n_c(T)>n^{*}(T)$ where $\Phi_{RS,q_>}(n)= f_{RS}$ and the RS solution $q_>(n)$ becomes the physical solution for all $n>n_c(T)$. 
Summarizing the behaviour of $n \Phi(n)$
as a function of $n$ in the high temperature phase is:
\beqa
n\, \Phi(n) & = & n f_{RS}\ \ {\rm for} \ n<n_c 
\nonumber
\\
n\, \Phi(n) & = & n_c f_{RS}+f_{RS,q_c}\Delta n+O(\Delta n)^2\ \ {\rm for} \ n>n_c 
\nonumber
\eeqa
where $\Delta n\equiv n-n_c$ and we have omitted the dependence on the temperature of $f_{RS}$, $n_c$ and $f_{RS,q_c} \equiv \partial n \Phi/\partial n|_{n=n_c^+}$.
At $n=n_c$ the order parameter jumps from zero to $q_c \equiv q_>(n_c)$, the free energy turns out to be discontinuous too:  $f_{RS,q_c}<f_{RS}$.
Thus as a function of $n$  $n \Phi(n)$ has a first order transition at $n=n_c(T)$, see fig. (\ref{dAT}). 
This peculiar behaviour of $\Phi(n)$ reflects itself in the following structure of $L(f)$ as follows from eqs. (\ref{prima},\ref{prima2},\ref{prima3},\ref{prima4}):
\beqa
L(f) &  = & 0 \ {\rm for} \ f=f_{RS}
\nonumber
\\
L(f) &  = & -\infty \ {\rm for} \ f_{RS,q_c}<f \neq f_{RS}
\nonumber
\\
L(f) & = &  -\beta n_c (f_{RS}-f_{RS,q_c})+\beta n_c \Delta f+O(\Delta f)^2  \ {\rm for} \ \Delta f \leq 0
\nonumber
\eeqa
where $\Delta f \equiv f-f_{RS,q_c}$.

To understand this double-peak behaviour of $L(f)$ it can be useful to think of the Random-Energy-Model (REM) \cite{REM}. In the typical sample the number of energy levels with energy $E$ is proportional to $\exp(N \ln 2-N E^2)$ and there are no levels outside the band $(-\sqrt{\ln 2},\sqrt{\ln 2})$. In the high-temperature phase the energy of the typical sample is given by the point where the derivative of $(\ln 2-E^2)$ is equal to $\beta$, {\it i.e.} $E=-\beta/2$ and the critical temperature is given by $\beta_c=2 \sqrt{\ln 2}$. The free energy is given by $F=-\beta/4-\ln 2/\beta$. Now in order to reduce the energy of such a sample at fixed $\beta<\beta_c$ one should modify the structure of the energy levels. One can see that any modification of the {\it global} shape of the distribution of the $N$ energy levels has a prohibitive price with a probability $O(\exp[-2^N])$, nevertheless one could instead pull a level out of the band with a cost in probability $O(\exp[N])$. Normally the energy of the lowest level is $-\sqrt{\ln 2}$ which is larger than the free energy of the levels with energy $-\beta/2$. Thus a small modification of the lowest level will not have any effect on the total free energy that will be still dominated by the levels with $E=-\beta/2$.
Only when the energy of the lowest level becomes smaller than $-\beta/4-\ln 2/\beta$ the thermodynamic of the sample is dominated by the lowest state and the energy jumps abruptly from $E=-\beta/2$ to a lower value $E=-\beta/4-\ln 2/\beta$.  

Coming back to the SK model we observe the following behaviour of $n_c(T)$ and $q_c(T)$
approaching the critical temperature:
\beq
\lim_{T \rightarrow 1}n_c(T)  = \lim_{T \rightarrow 1}n^*(T)=2
\eeq
\beq
\lim_{T \rightarrow 1}q_c(T) =  \lim_{T \rightarrow 1}q_<(T)=0 
\eeq
Therefore at the point $(T=T_c,n=2)$ the discontinuity in the free energy vanishes and it represents the end-point of a the line $n_c(T)$ of first order phase transitions.
More precisely near the critical point $T=T_c$ and $n=2$ we have at leading order in $\tau=T_c-T<0$:
\beqa
n^*(T) & \simeq  & 2+\left(-{8 \over 3} \tau \right)^{1/2}
\nonumber
\\
\Phi_{RS,q_<}(n^*) =  \Phi_{RS,q_>}(n^*) & \simeq  &  f_{RS}+{\tau^2 \over 2}
\nonumber
\\
n_c(T) & \simeq  & 2+ \left(-3 \tau \right)^{1/2}
\nonumber
\\
q_c(T) & \simeq  & 2 \left(-3 \tau \right)^{1/2}
\nonumber
\\
f_{RS,q_c}(T)-f_{RS}(T) & \simeq  & - 8 \sqrt{3} \left(- \tau \right)^{3/2}
\nonumber
\eeqa

\subsection{At the Critical Temperature}
On the line $T=1$ we have $\Phi(n,T_c)=f_{RS,q=0}(T_c)=-1/4+\ln 2$ for $n<2$ while for positive  $\Delta n \equiv n-2$ the solution is still RS but with a non zero value of $q$; at leading orders we have:
\beqa
q & = & 3 \Delta n +O(\Delta n^{2})
\\
\Phi(n,1) & = & f_{RS,q=0}-{9 \over 8} \Delta n^4+O(\Delta n^5)
\eeqa
Note that at the critical temperature the range where the sample complexity is finite touches $f_{RS}$, and $L(\Delta f)=2 \Delta f+O(\Delta f^2)$. This behaviour is interesting in connection with the problem of the small deviations of the free energy as we will discuss below. 

\subsection{The low-Temperature Replica-Symmetric Region}
On the straight line that connects the point $(T=1,n=0)$ and $(T=1,n=2)$ the potential $\Phi(n)$ has  also a phase transition, see fig. (\ref{dAT}).
On the right of this line we have $q=0$ and $\Phi(n,T)=f_{RS,q=0}(T)=-\beta/4+\ln 2 /\beta$. On the left the solution is still RS (except for $n=0$, see below) and the parameter $q$ has a continuous transition. At the leading order in $\tau=T_c-T$ we have (for $n<2$ and $0<\tau \ll 2-n$):
\beqa
q(n,T) & = & {2 \over 2-n}\tau + O(\tau^2)
\\
\Phi(n,T) & = & f_{RS,q=0}+{2 (1-n)\over 3 (n-2)^2} \tau^3+O(\tau^4)
\label{phis3}
\eeqa   
Note that the physical $\Phi(n,T)$ is equal to $f_{RS,q=0}(T)=-\beta/4-\ln 2 /\beta$ at {\it any} temperature for $n=1$, while it is larger for $n<1$ and smaller for $n>1$.

The above expansion in powers of $\tau$ breaks down at $n=2$. On the $n=2$ line the RS solution satisfies the exact equation:
\beq
q= \tanh \beta q
\eeq
this leads to the following behaviour at leading order in $\tau$
\beqa
q & = & \sqrt{6 \tau}+O(\tau^{3/2})
\\
\Phi(2,T) & = & f_{RS,q=0}-{3 \over 2} \tau^2+O(\tau^3)
\eeqa

On the other hand near $n=0$ the RS solution is inconsistent, indeed for convexity the function $n \Phi(n)$ must have a negative second derivative with respect to $n$ but this condition fails at any $\tau$ for sufficiently small values of $n$.
According to eq. (\ref{phis3}) we have for small positive $\tau$:
\beq
{ \partial^2 n \Phi(n) \over \partial n^2}=-{4 n \tau^3 \over (n-2)^4}+O(\tau^4)
\eeq 
thus this quantity goes to zero for small $n$ and we have to take care of the $O(\tau^4)$ term.
Taking into account the $O(\tau^4)$ term and expanding in powers of $n$ we have:
\beq
\Phi_{RS}(n,\tau)-\Phi_{RS}(0,\tau)=-{n^2 \tau^3 \over 24}+{n \tau^4 \over 12}+O(\tau,n,5)
\eeq
from the above equation we see that for $n<{2 \tau/3}+O(\tau)^2$ the second derivative of the $n \Phi(n)$ would be positive and the solution inconsistent.
Thus in the $(\beta n,T)$ plane there is a line $n_{conv}(\tau)={2 \tau/3}+O(\tau)^2$ below which the RS solution is inconsistent for convexity reasons and cannot be the correct one. In the following we will see that actually the RS solution becomes unstable and should be discarded below a line $n_{dAT}(\tau)={4 \tau/3}+O(\tau)^2$ which is above the line $n_{conv}(T)$. 

\begin{figure}[htb]
\begin{center}
\epsfig{file=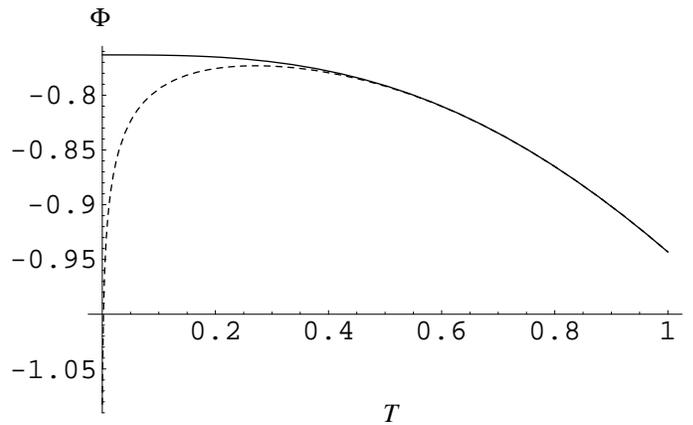,width=9cm}
\caption{The potential $\Phi(n,T)$ vs. temperature at equilibrium ($n=0$) (solid) \cite{CR} and on the dAT line ($n=n_{dAT}(T)$) (dashed), for small $\tau=T_c-T$ the difference  is $\Phi(0,T)-\Phi(n_{dAT}(T),T)=\tau^5/135+O(\tau^6)$. The Potential on the dAT line diverges as $-\beta n_{dAT}(T)/4-\ln 2/\beta n_{dAT}$ at low temperatures. }
\label{fdat}
\end{center}\end{figure}

In the low temperature phase the replica solution is unstable at small values of $n$ as the replicon eigenvalue becomes negative \cite{MPV}. 
Similarly to the stability of the RS solution in the magnetic-field/Temperature plane, in the $(n,T)$ plane the region of stability is above the deAlmeida-Thouless (dAT) line that is specified by the condition:
\beq
T^2={ 
\int e^{-{y^2\over 2 q}}(\cosh \beta y)^n (1-\tanh^2 \beta y)^2 \, dy
\over
\int e^{-{y^2\over 2 q}}(\cosh \beta y)^n \, dy
}
\eeq
For small $\tau=1-T$ the value of $n$ on the dAT line is 
\beq
n_{dAT}(T)= {4 \over 3} \tau +O(\tau^2) 
\eeq
while in the zero temperature limit $q$ goes to unity and we have:
\beq
n_{dAT}(T)=T\sqrt{-2 \ln \left[ 3 \left( {\pi \over 2} \right)^{1/2} T\right]}
\eeq 
Note that $n_{dAT}$ vanishes in the zero-temperature limit but in the $(T, n \beta)$ plane the dAT line never touches the $T=0$ line and the sample complexity $L(e)$ at $T=0$ is always in the RSB phase, see fig (\ref{dAT}).
On the other hand this show that at any {\it fixed} $n>0$ the system exit the RSB phase at low enough temperature and the solution is always RS at zero temperature. 
In figure (\ref{fdat}) we plot the potential $\Phi(n,T)$ on the dAT line, it goes to minus infinity at low temperature as:
\beq
\Phi(n_{dAT}(T),T)=-{\beta \, n_{dAT}(T) \over 4} -{\ln 2 \over \beta \, n_{dAT}(T)}
\eeq  
Note that the second term gives a vanishing correction that can be rather large at finite temperature. 

\subsection{The Large $n$ Limit}

At any finite temperature  $\Phi(n)$ is described by the RS solution at large values of $n$. Both above and below the critical temperature, the behaviour of $\Phi(n)$ for large values of $n$ is $\Phi(n)=-\beta n/4-\ln 2 /(\beta n)+O(e^{-2\beta n})$. This leads to $L(f)=-f^2+\ln 2+o(1)$ for large negative $f$, note that this is the same behaviour of the Random-Energy-Model (REM)\cite{REM}.

\section{The Replica-Symmetry-Breaking Phase}
\label{RSB}

Below the dAT line $n<n_{dAT}(T)$ we must break the replica symmetry.  As we can see in fig. (\ref{fdat}) $\Phi(n)$ on the dAT line is smaller than the most likely free energy ($f_{typ}$) that is $\Phi(n)$ at $n=0$ at the same temperature, in particular $\Phi(n_{dAT})$ diverges as $-\beta n_{dAT}(T)/4-\ln 2/\beta n_{dAT}$ at low
temperatures while for small $\tau=T_c-T$ the difference is $\Phi(0,T)-\Phi(n_{dAT}(T),T)=\tau^5/135+O(\tau^6)$.  On the other hand being a convex function $\Phi(n)$ must be continuous,
therefore we must look for a free energy that shows some dependence on $n$
also below the dAT line and the one suggested by Kondor is the most natural one.

We recall that in Kondor's approach  for $n<n_{dAT}(T)<1$  one introduces a function $q(x)$ defined
for $n\le x\le 1$ that describes the breaking of replica symmetry in the low temperature phase. A functional $F_{n}[q]$ is obtained such that
$\Phi(n)=\max_{q}F_{n}[q]$.  The function $q(x)$ that maximizes $F_{n}[q]$ can be found by solving the
stationarity equation $\delta F / \delta q(x) = 0$.  This generalizes the standard approach that is proved
to give the correct value of $\Phi(n)$ in the $n \rightarrow 0$ limit. 

The form of the free energy functional is the usual one \cite{MPV}, the only difference being that all functions are defined
in the interval $n\le x \le 1$: 
\begin{displaymath}
F_n[q(x)]  \equiv  -{\beta \over 4}\left(1- 2 q(1)+\int_n^1  q^2(x)\, dx\right)
+
\end{displaymath}
\begin{displaymath}
 - {1 \over \beta n}\ln \int_{-\infty}^{+\infty}{dy \over \sqrt{2 \pi  q(n)}}\exp \left(-{(y-h)^2\over 2 q(n)}\right)\exp(\beta n f(n,y))
\end{displaymath}
the function $f(x,y)$ obeys the following equation:
\beq
\dot{f}=-{\dot{q} \over 2}(f''+\beta x (f')^2)
\label{eqf}
\eeq
where dots and primes mean respectively derivatives with respect to $x$ and $y$. The initial condition is
\beq
f(1,y)={1 \over \beta}\log 2 \cosh \beta y
\label{inif}
\eeq
The above functional has to be extremized with respect to the function $q(x)$. 
A set of variational equations can be obtained introducing Lagrange multiplier $P(x,y)$ to enforce equations (\ref{eqf}) and (\ref{inif})  \cite{SD,CR} , the resulting equations are:
\beq
q(x)=\int_{-\infty}^{\infty}P(x,y) m^2(x,y) dy
\eeq
\beq
m=f'\, ; \ \ \ \dot{m}=-{\dot{q}\over 2}(m''+2 x \beta \,m \,m' )
\eeq
\beq
\dot{P}={\dot{q}\over 2}(P''-2 x \beta \,(m \,P)' )
\eeq
These are the same equations of the standard $n\rightarrow 0$ case, the only difference is in the initial condition for $P(x,y)$ that reads:
\beq
P(0,y)= c \, \exp \left[ -{(y-h)^2 \over 2 \, q(n)}+ \beta \, n \, f(n,y)  \right]
\eeq
where $c$ is a normalization constant in order to have $\int P(x,y) dy=1$. Since $\Phi(n)$ is extremized with respect to $q(x)$, the conjugate variable $f$ can be obtained as the derivative of the $n \Phi(n,q(x))$ evaluated at the saddle point:
\beq
f=-{\beta \over 4}\left(1- 2 q(1)+\int_n^1  q^2(x) \, dx -n q^2(n) \right)-\langle f(n,y) \rangle
\eeq  
where the square brackets represent average with respect to the measure $ d \mu = \exp (-(y-h)^2 / 2 q(n) + \beta n f(n,y))$.

We have solved the RSB equations and computed $q(x)$ and $\Phi(n,\beta)$ as power series of $n$ and $\tau=1-T$ \cite{CR}; the power series of $\Phi(n)$ up to 18th order is reported in the appendix.
Kondor originally used the so-called truncated model valid near the critical temperature and he found that $\Phi(n)=f_{eq}-9 n^5/5120$ \cite{Kon1}. At all orders considered we have confirmed that the lowest power of $n$ in the expansion of $\Phi(n)$ is $n^5$ and that there is also no $n^6$ term. 
We have verified  by an expansion in
powers of $n$ at {\it fixed} temperature that {\it the first term in   $\Phi(n)$ is of $O(n^5)$  at all temperatures} as follows from an  analytic argument presented in the appendices. This result is also related to the behaviour of the free energy functional with increasing number of RSB steps \cite{Opper}.  
An alternative argument can be done  using the expansion of the replicated free energy functional $F[Q_{ab}]$ \cite{MPV} in
powers of the $n \times n$ matrix $Q_{ab}$, at least for powers less than 10, where one can use the
explicite form of the terms. The expressions become more complex when
the power of Q become larger or equal to 10.

For negative $n$ the saddle point of the $\Phi(n)$ is the standard $q(x)$ corresponding to $n=0$, thus $\Phi(n)=f_{eq}$ for $n<0$ \cite{DFM}.
The corresponding sample complexity as a function of $\Delta f=f-f_{typ}$  reads:
\beqa
L(f) &  = & -\infty \  \ \ \ \ \ \ \ \ \ \ \ \ \ \ \ \ \ \ \ \ \ \ \ \ \ \ \ \ \ \   {\rm for}\  \Delta f>0
\nonumber
\\ 
L(f) &  = & a_{6/5} |\Delta f|^{6 / 5} +O(|\Delta f|^{8 / 5})  \ \ \ {\rm for} \ \Delta f \leq 0
\nonumber
\eeqa
Where $a_{6/5}=-5\beta  |c_5|^{-{1 / 5}} 6^{-{6 / 5}}$ and $c_5$ is the coefficient of $n^5$ in the expansion of $\Phi(n)$.
The function $q(x)$ has a small  plateau from $n$ to some value $x_c$. For $x_c<x<1$ $q(x)$ has the usual shape, more precisely deviations of $q(x)$ from $q_{free}(x)$, {\it i.e.} the solution corresponding to $n=0$, are $O(n^5)$ in this region. We note that $q(x)$ is always continuous at the end-point of the first plateau $x_c$.
At the leading order in $n$ we have:
\beqa
q(x)  & = & {3\over 2}n \dot{q}(0) +O(n^2) \ \ \ \ \ \  {\rm for}\  n<x<x_c \equiv{3 \over 2}n+O(n^2)
\nonumber
\\
q(x)  & =  & q_{free}(x)+O(n^5)  \ \ \ \ \ {\rm for} \ x_c<x<1
\nonumber
\eeqa 
where $\dot{q}(0)$ is the derivative of $q_{free}(x)$ in $x=0$.
Note that the above expression are valid at {\it all} temperatures since they have been obtained from the {\it fixed} temperature expansion in powers of $n$ reported in the appendices.
 We have also verified directly these features of $q(x)$ 
at the order to which we computed the expansion in $n$ and $\tau$.

It is interesting to note that from the third order on, all derivatives of $\Phi(n)$ (with respect to $n$,$T$ and both) are discontinuous on the dAT line {\it i.e.} the transition is third order. This is the same behaviour of the free energy on the dAT line in the $(h,T)$ plane \cite{CRT}. 

When $\beta \rightarrow \infty$ the complexity $L(f)$ goes to a well-defined limit therefore from eq. (\ref{prima}) $\Phi(n)$ is actually a function of $\beta n$ \footnote{This result is related to the scaling in the  $T \rightarrow 0$ limit of the function $q(x)$: $q(x,\beta)\approx \hat q(\beta x)$.} , as a consequence the coefficient $c_a$ of $n^a$ in the power series of $\Phi(n)$ diverges as $\beta^a$ in the zero temperature limit. 

\begin{figure}[htb]
\begin{center}
\epsfig{file=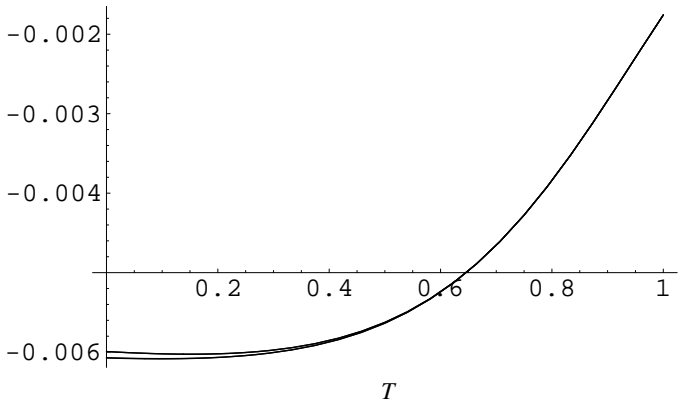,width=9cm}
\caption{Four Pad\'e approximants ( $P(3,3)$, $P(3,4)$, $P(4,3)$, $P(4,4)$ ) of $c_5 \beta^{-5}$ as a function of the temperature, $c_5 \beta^{-5}$ is equal to $-9/5120$ at $T=1$.}
\label{p}
\end{center}\end{figure}

The series in power of $\tau$ of $c_5$ (the $n^5$ coefficient in $\Phi(n)$) can be used to obtain its behaviour in the whole low temperature phase provided one uses the information that $c_5 \sim \beta^5$ in the zero-temperature limit. In figure \ref{p} we plot various Pad\'e approximants obtained from the series of $c_5 \beta^{-5}$.
From the Pad\'e approximants of $c_5 \beta^{-5}$ and $c_7 \beta^{-7}$ we estimate $c_5 \simeq -0.0060(2) \, \beta^5 $ near $T=0$ and $c_7 \simeq -0.0150(5) \beta^7$ in the SK model. 
Alternatively one can use the following {\it exact} relationship valid at all temperatures (see appendix \ref{xplat}):
\beq
c_5=-{9 \over 640} \beta^5 (T \dot{q}(0))^3
\eeq
where $\dot{q}(0)$ is the derivative in $x=0$ of the usual function $q(x)$ for $n \rightarrow 0$.
The zero-temperature limit of $T \dot{q}(0)$ is finite and was obtained resumming its expansion in powers of $\tau$ as $.743(2)$ in \cite{CR}. Recently a more precise estimate $.743368$  has been obtained working directly at zero temperature \cite{Opper}, this gives $\lim_{\beta \rightarrow \infty} c_5 \beta^{-5} = -0.0057766 $. 

The zero temperature complexity for negative $\Delta e$ then reads:
\begin{displaymath}
L(\Delta e)=-1.63250 \, |\Delta e|^{6/5}+3.1(1)\, |\Delta e|^{8/5}+O(\Delta e^{8/5})
\label{DSx}
\end{displaymath}
The second term however yields a big correction to the first one,  indeed: i) the exponents of the series grow slowly (as $(6+i)/5$, $i=0,2,3,\dots$, note that there is no $n^6$ term in $\Phi(n)$) and ii)  the coefficients of the series grow quickly with order, actually we expect the series to be asymptotic as is usually the case in this context \cite{CR}.
Therefore in order to have a good control on $L(\Delta e)$ we have adopted a method previously used in \cite{CR} to obtain $q(x,\tau)$ from its series in powers of $x$ and $\tau$. We have transformed the series of $L(\Delta f)$ in powers of $\Delta f$ and $\tau$ in a power series of just $\tau$ by setting $\Delta f=({2 \over 45}s^5+{1\over 4}\tau^7)c$ with $c$ a parameter in the range $[0,1]$.  The corresponding series in powers of $\tau$ were resummed for any given  $c$ through Pad\'e approximants obtaining the curve $L(\Delta e)$ in parametric form.

\begin{figure}[htb]
\begin{center}
\epsfig{file=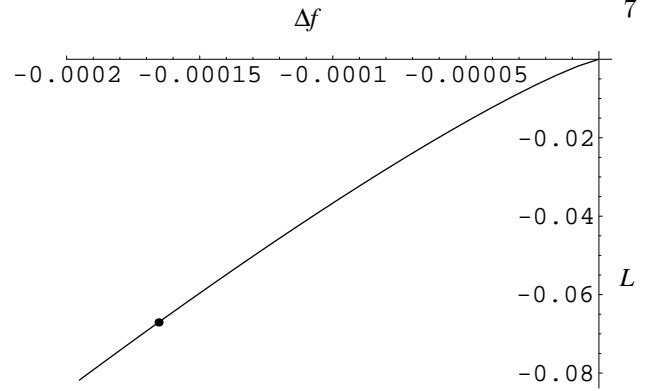,width=8cm}
\caption{Sample complexity $L(\Delta f)$ at $T=.7$, obtained through an $(8,5)$ Pad\'e approximant to the parametric power series in $\tau$. The dot marks the RSB-RS transition at $\Delta f=-1.64 \times 10^{-4}$.}
\label{p8507}
\end{center}\end{figure}

In fig. (\ref{p8507}) we have plotted the function $L(\Delta f)$ at temperature $T=.7$ obtained by resumming the series of $L(\Delta f(\tau,c)))/\tau^6$ by means of a Pad\'e approximant of order $(8,5)$ (we have used the series of $\Phi(n)$ to $18th$ order reported in the appendix).

In figure (\ref{numeric}) we plot the sample complexity $L(\Delta e)$  at zero temperature. 
In the range of energy differences considered the deviations from the values yielded by $L(\Delta e)=-1.63250 \, |\Delta e|^{6/5}$ are no larger than $1\%$, this support the goodness of both estimates since they were obtained by different resummation schemes.

By resumming the series of $\Phi(n)$, (see below) we have been able to obtain the sample complexity in the whole low-temperature phase and for finite $L(f)$, in figure \ref{numeric} we compare the sample complexity with the numerical data at zero temperature of Ref. \cite{CMPP} finding a very good agreement.
For each system size $N$ we have plotted $L_N=\ln(P(\Delta e_N)/N^{5/6})/N$ with $\Delta e_N=e-e_N$ (the average energy at size $N$), we have used this definition so that $L$ goes to a constant for $\Delta e_N=0$.  
The errors on $L_N$ have been computed through error propagation and single events have been discarded to reduce the error.  The quantitative agreement of the numerical with the theory is quite good.

\begin{figure}[htb]
\begin{center}
\epsfig{file=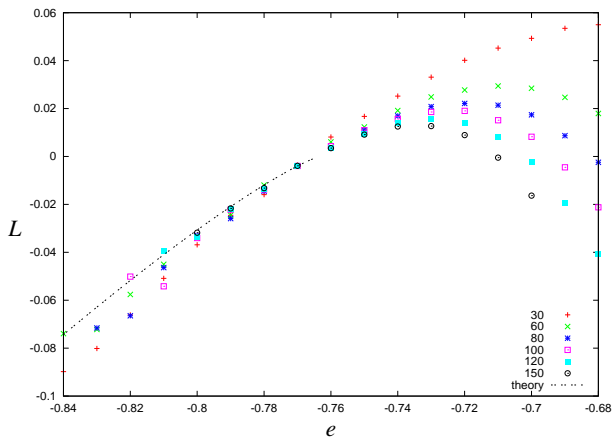,width=8cm}
\caption{Plot of the numerical complexity (from Ref. \cite{CMPP}) as a function 
of the absolute value of the energy for different sample sizes at 
zero temperature. The data have been shifted vertically by an 
amount $\Delta_N$ so that the complexity vanishes at the typical 
energy $E_{typ}=-.7633$. }
\label{numeric2}
\end{center}\end{figure}

In figure \ref{numeric2} we have also plotted the numerical complexity (from Ref. \cite{CMPP}) as a function 
of the absolute value of the energy for different sample sizes at 
zero temperature. The data have been shifted vertically by an 
amount $\Delta_N$ so that the complexity vanishes at the typical 
energy $E_{typ}=-.7633$. Since this certainly holds in the 
thermodynamic limit $\Delta_N$ goes to zero at large $N$. Note 
that for $E<E_{typ}$ the numerical data approach the theoretical 
prediction from below, this rules out the alternative prediction 
of Ref. \cite{AM} that yields $L(\Delta e)=-\infty$.

\begin{figure}[htb]
\begin{center}
\epsfig{file=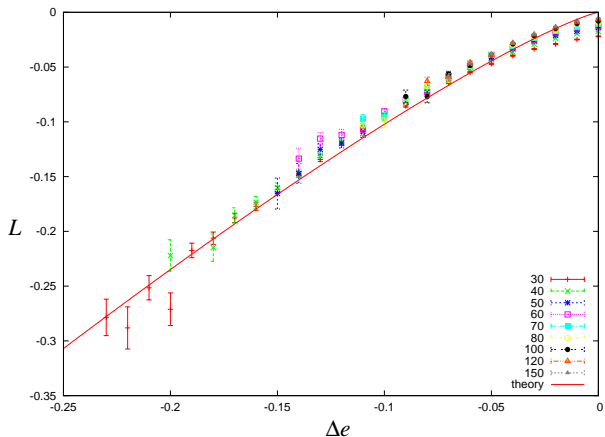,width=8cm}
\caption{Comparison between the numerical and analytical sample complexity at zero temperature, see text. The data are those of Ref. \cite{CMPP}. The sample complexity was obtained through an $(8,5)$ Pad\'e approximant to the parametric power series in $\tau$, the deviations from the expression $L(\Delta e)=-1.63250 \, |\Delta e|^{6/5}$ are less than $1\%$ in this range of energy differences.}
\label{numeric}
\end{center}\end{figure}

\section{On The Mean-Field Spherical Spin-Glass Model}
\label{SPHERICAL}
The RS expression of $\Phi(n,T)$ of the spherical model is \cite{KTJ}:
\begin{displaymath}
\Phi(n,T)[z,q]  =  -{\beta \over 4}-{z \over \beta}-{1-n \over 4}\beta q^2+
\end{displaymath}
\beq
+{1 \over 2 \beta}\left( \ln[z+{\beta^2 \over 2}q(1-n)]+(n-1)\ln[z+{\beta^2 \over 2}q]\right)
\eeq
The above expression has to be extremized with respect the parameters $q$ and $z$.
The phase diagram of the $\Phi(n,T)$ in the $(T,\beta n)$  plane is qualitatively similar to that of the SK model, see fig. (\ref{dAT}) , except for the absence of the dAT line.
In particular above $T_c=1$ the solution is paramagnetic ($q=0$) for small $n$ and does not depend on $n$,
\beq
 \Phi(n,\beta)=-{1 \over 2 \beta}-{\beta \over 4}-{\ln 2 
\over 2 \beta}
\label{sphq0}
\eeq
at some temperature dependent value $n=n_c(T)$ there is a first order transition. The line $n_c(T)$ ends on the point $(T=1,n=2)$, around that point it goes as $n_c(\tau)\simeq2+3 \sqrt{-\tau}$ for negative $\tau=T_c-T$. The straight line from the point $(T=1,n=0)$ to the point $(T=1,n=2)$ divides the paramagnetic from the spin-glass ($q \neq 0$), at variance with the $n_c(T)$ line, the parameter $q$ varies continuously upon crossing this line.
At all temperature the physical value of $\Phi(n,T)$ is smaller than expression (\ref{sphq0}) for $n>1$ while  it is larger or equal to it for $n \leq 1$.

Below the critical temperature only the spin-glass phase is present. The main difference with respect to the SK model is the absence of the dAT line in the phase diagram, meaning that the RS solution remains correct in the limit $n \rightarrow 0$ \cite{KTJ}.
The expression of $\Phi(n,T)$ for $T<1$ at small values of $n$ is:
\beq
\Phi(n,\beta)={1 \over 4 \beta}-1-{\ln [\beta/2] 
\over 2 \beta}-{(\beta-1)^3\over 24 \beta}n^2+O(n^3)
\eeq
Note that again the linear term in $n$ is missing and the sample complexity is non-Gaussian:
\beqa
L(f) & = & -\infty \ \ \ \ \ \ \ \ \ \ \ \ \ \ \ \ \ \ \ \ \ \ \ \ \ \ \ \ \ {\rm for}\ \Delta f>0
\nonumber
\\
L(f) & = &  -{4 \sqrt{2} \over 3}|\Delta f|^{3/2}\left( \beta \over \beta-1\right)^{3/2}   \ \ {\rm for}\ \Delta f \leq 0
\label{large}
\eeqa
At zero temperature the energy of the model is equal to minus the largest eigenvalue of a Gaussian random matrix
\beq
e=-{\lambda_{max} \over \sqrt{2 N}}
\label{eqe}
\eeq 
The small deviation distribution of the largest eigenvalue of a Gaussian random matrix is given by the Tracy-Widom law $F_1(x)$ \cite{TW} in terms of the rescaled variable $x=\sqrt{2}(\lambda_{max}-\sqrt{2 N})N^{1/6}$.
The behaviour of $F_1(x)$ for $x \rightarrow +\infty$ is:
\beq
\ln F_1(x) \sim -{2 \over 3}x^{3/2}=-{4 \sqrt{2} \over 3}|\Delta e|^{3/2} N
\label{small}
\eeq
where we have used   $x=2 N^{2/3}|\Delta e|$ as follows from eq. (\ref{eqe}). Thus eq. (\ref{large}) and (\ref{small}) give the same prediction and there is perfect matching between small and large deviations. A similar matching has been also observed for positive deviations $\Delta e$ whose probability scales as $\exp[O(N^2)]$, \cite{DM}.

\section{On Small Deviations}
\label{small}

In this section we discuss the connection between small deviations and large deviations of the free energy.
The function $\Phi(n)$ is the natural object to describe the large deviations of $f$ from its typical value.
In \cite{CPSV} it was argued that it provides also information on the small deviations of the free energy arguing that they scale as $N^{-5/6}$.

The probability distribution of the free energy {\it per spin} $P_N(f)$ is concentrated near the typical free energy $f_{typ}$ in the large $N$ limit. The small deviations corresponds to values of the free energy difference that have a finite probability $P_N(f)=O(1)$ to be observed in the thermodynamic limit, that is a region near $f_{typ}$ that shrinks to zero in the thermodynamic limit.
On the other hand the large deviations corresponds to the exponentially small tails of $P_N(f)$ corresponding to $O(1)$ values of the free energy difference.
Thus in principle small and large deviations are fairly different objects and it may seems strange that one can determine the scaling of the peak from a large deviation calculation.

Typically the probability of the small deviations scales as $\lim_{N \rightarrow \infty} P_N(f)=p((f-f_N)/N^{-a})$ where $p(x)$ is a scaling function that does not depend on $N$, $a$ is some positive exponent and $f_N$ is the $N$-dependent mean value of the free energy that converges to $f_{typ}$ as $f_N=f_{typ}+N^{-b}$ for some positive $b$.

The argument that connects large and small deviations is not rigorous and relies on the assumption that there is smooth matching between the behaviour of the peak of $P_N(f)$ and the left tail corresponding to positive values of $n$. Under this assumption one argues that the region of the peak corresponds to values of the free energy difference  $(f_{typ}-f)$ such that the large deviation expression (\ref{LP}) is finite, this happens for $(f_{typ}-f)=O(N^{-5/6})$ and leads to the aforementioned prediction $a=5/6$. In other words the matching argument corresponds to the assumption that the function $p(x)$ provides a good description of the distribution of the free energies up to free energies differences $\Delta f=O(1)$, {\it i.e.} far beyond its natural range of validity $\Delta f=O(N^{-a})$.   

It is interesting to note that  in order to characterize the small deviations one should take the $n \rightarrow 0$ limit first and then the $N \rightarrow \infty$ limit while the two limits have to be inverted to obtain the large deviations.

In the following we discuss what kind of quantitative information can be extracted from the thermodynamic limit of $\Phi(n)$ under the assumption that there is a smooth matching between small and large deviations.

The starting observation is that at any {\it finite} value of $N$, $n \Phi(n)$ is the generating function of the cumulants  of the distribution of $F$. If in the thermodynamic limit $\Phi(n)=f_{typ}+c_a \,n^a N^{-b}$ one would consider the variable $ x=(F-F_N)/N^{(1-b)/(a+1)}$, where $F_N$ is the average free energy at size $N$, and claim that the $a-th$ cumulant of its distribution function $f(x)$ is finite while all higher cumulants are zero. In particular in the high temperature phase we have $\Phi(n)=f_{typ}+c_a \,n N^{-1}$ and we would say that the fluctuations of $F$ around its average value $F_N$ are normal with finite variance because all cumulants higher than the second vanish.

Extending this argument to the low temperature case one could say that all the cumulants greater than the sixth of the variable $ x=(F-F_N)/N^{1/6}$ vanish.
This conclusion however is wrong because the function $\Phi(n)$ in the low temperature phase has a first order phase transition at $n=0$ and the two limit $N \rightarrow \infty$ and $n \rightarrow 0$ cannot be exchanged in computing derivatives of $\Phi(n)$. The only exception is the zero-th derivative $\Phi(0)$ (the average free energy).
This can be also understood noticing that by fixing $n$ and taking the limit $N \rightarrow \infty$ the actual value of $x$ goes to infinity,  therefore there is in principle no way to get information on the small-$x$ region once the thermodynamic limit has been taken. However under the assumption of smooth matching between small and large deviations it is also natural to assume that 
 the small deviations of the free energy  behave for large negative $x$ as:
\beq
p(x) \sim \exp ( a_{6/5} |x|^{6 / 5} ) 
\eeq
In this context it is instructive to consider the REM at zero temperature \cite{REM}.  The complexity as a function of $\Delta e \equiv e-\sqrt{\ln 2}$ behaves as $L (\Delta e)=-\beta_c\Delta e$ for $\Delta e <0$  (with $\beta_c=2 \sqrt{\ln 2}$) while  $L =-\infty$ for positive $\Delta e$. 
Using the matching argument between small and large deviations we would conclude that the extensive energy has finite variance and that
 the behaviour of the rescaled variable $x=N(e-e_N)$ for $x$ negative and large is $\exp(- \beta_c x)$, this prediction is consistent with the known fact that the small deviations obey the Gumbel law. However in order to recover the full Gumbel distribution of small deviations  we should take the $n \rightarrow 0$ limit first. Note also that the deviations of $E$ from its mean $E_N$ (which is $O(1)$ in this case) has nothing to do with the deviations of $E_N$ from its thermodynamic limit which is $O(\ln N)$ in the REM \cite{REM}.
 
Another example of matching between small and large deviations is provided by the spherical model as discussed in the previous section. 

On the other hand the very same SK model at the critical temperature seems to provide an example of the failure of the matching argument. Indeed above the critical temperature the sample complexity is $-\infty$ for $\Delta f \neq 0$ and has no role in the finite-size fluctuations of the free energy which instead are controlled by corrections to the $q=0$ solutions \cite{GRS,CPSV}. The  free energy $F$ has a finite variance that diverges for $T \rightarrow T_c$. Therefore it is expected that the free energy variance diverges with $N$ at the critical temperature. Indeed extending the  computation of Ref. \cite{GRS} Aspelmeier has recently argued that the variance diverges logarithmically with $N$ \cite{ACHAOS}. On the other hand at the critical temperature the range where the sample complexity is finite touches $f_{RS}$, and $L(\Delta f)=2 \Delta f+O(\Delta f^2)$; therefore applying the matching argument one would wrongly conclude that the behaviour of the distribution of $F$ is $\exp(2 F)$ for large negative $F$ and that its variance remains finite.
However we believe that the matching argument is correct for the SK model below the critical temperature and that its failure at the critical temperature can be explained noticing that there is a phase transition in the $(n,T)$ plane on the straight line connecting the point $(0,1)$ and $(2,0)$ and all the eigenvalues of $\Phi[Q^{ab}]$ vanish on this line. 

\section{Conclusions}
\label{C}

We have computed the function $\Phi(n,T)$ of the SK model using the hierarchical ansatz and discussed its behaviour in detail both above and below the DeAlmeida-Thouless line in the $(n,T)$ plane.
In particular in the low-temperature phase we have confirmed at all orders 
Kondor's early result \cite{Kon1} on the  $O(n^5)$ scaling.
The analytical argument provides an exact relationship between the coefficient of the $n^5$ term and the $x=0$ derivative of the the standard function $q(x)$. We note that the same approach provides a similar exact relationship between $\dot{q}(0)$ and the $O(h^{10/3})$ term in the equilibrium free energy in presence of a magnetic field $h$.
By resumming the series we have been able to obtain for the first time the sample complexity at zero temperature. 
Existing numerical data display a remarkable agreement with this prediction.  
We mention that the presence of a magnetic field reintroduces a $O(n^2)$ dependence in $n \Phi(n)$ leading back to Gaussian fluctuations of the free energy.

\appendix

\section{Power Series of $\Phi(n)$}
\label{appa}
In this appendix we report  the power series of $\Phi(n)$ of the SK model in the low temperature phase up to the 18th order in $n$ and $\tau=1-T$. At all order in $\tau$ the smallest power of $n$ is $n^5$ and there is no $n^6$ term.
\begin{widetext}
\beqa
\Phi(n) & = & - \frac{1}{4} -\ln 2 -\frac{\tau}{4}+\tau \ln 2 - \frac{{\tau}^2}{4} - \frac{{\tau}^3}{12} + \frac{{\tau}^4}{24} - \frac{{\tau}^5}{120} + 
  \frac{3\,{\tau}^6}{20} - \frac{79\,{\tau}^7}{140} + \frac{1679\,{\tau}^8}{560} - \frac{13679\,{\tau}^9}{720} + 
  \frac{1728361\,{\tau}^{10}}{12600} +
  \nonumber
  \\
& & -\frac{19214684\,{\tau}^{11}}{17325} + \frac{2741593487\,{\tau}^{12}}{277200} - 
  \frac{3939806687\,{\tau}^{13}}{40950} + \frac{773933492429\,{\tau}^{14}}{764400} - \frac{86662083146207\,{\tau}^{15}}{7567560} +
\nonumber
\\
& & + 
  \frac{139738065304401461\,{\tau}^{16}}{1009008000} - \frac{45875375549246420713\,{\tau}^{17}}{25729704000} + 
  \frac{11276190176083149262457\,{\tau}^{18}}{463134672000} +  
  \nonumber
  \\
& n^5 & \left( - \frac{9}{5120}   - \frac{99\,\tau}{5120} - \frac{27\,{\tau}^2}{320} - \frac{279\,{\tau}^3}{1280} - 
  \frac{981\,{\tau}^4}{2560} - \frac{351\,{\tau}^5}{400} + \frac{2799\,{\tau}^6}{12800} - 
  \frac{344241\,{\tau}^7}{22400} + \frac{47010861\,{\tau}^8}{358400}+ \right.
  \nonumber
  \\
& &
\left. - \frac{36684189\,{\tau}^9}{25600} + 
  \frac{830566899\,{\tau}^{10}}{51200} - \frac{1928757352257\,{\tau}^{11}}{9856000} + 
  \frac{98506298782713\,{\tau}^{12}}{39424000} - \frac{8635947355938261\,{\tau}^{13}}{256256000}\right)+
\nonumber
\\
& n^7 & 
\left( \frac{81}{143360} - \frac{2673\,\tau}{143360} - \frac{7047\,{\tau}^2}{35840} - \frac{35559\,{\tau}^3}{35840} - 
  \frac{75573\,{\tau}^4}{35840} - \frac{1943757\,{\tau}^5}{179200} + \frac{7442847\,{\tau}^6}{179200}+\right.
  \nonumber
\\  
& & \left.
-  \frac{762113853\,{\tau}^7}{1254400} + \frac{17011569051\,{\tau}^8}{2508800} 
 - \frac{210723811119\,{\tau}^9}{2508800} 
  +\frac{13663823711841\,{\tau}^{10}}{12544000} - \frac{1027400967213903\,{\tau}^{11}}{68992000}\right)+
\nonumber
\\
& n^8 & \left(
\frac{243}{32768} + \frac{4131\,\tau}{32768} + \frac{15309\,{\tau}^2}{16384} + \frac{34263\,{\tau}^3}{8192} + 
  \frac{429381\,{\tau}^4}{32768} + \frac{2740311\,{\tau}^5}{81920} + \frac{11253573\,{\tau}^6}{163840} +\right.
\nonumber
\\
& & \left. + 
  \frac{107945217\,{\tau}^7}{573440}
 - \frac{669127959\,{\tau}^8}{4587520} + \frac{12126319893\,{\tau}^9}{2293760} - 
  \frac{183401224893\,{\tau}^{10}}{3276800}
\right)+
\nonumber
\\
& n^9 & \left( 
-\frac{60021}{5734400}   - \frac{1720683\,\tau}{5734400} - \frac{3703563\,{\tau}^2}{1433600} - 
  \frac{48430143\,{\tau}^3}{2867200} - \frac{213993819\,{\tau}^4}{5734400} - \frac{2813451327\,{\tau}^5}{7168000} +\right.
\nonumber
\\
& & \left.
+ 
  \frac{2765750427\,{\tau}^6}{1146880} - \frac{1836578874951\,{\tau}^7}{50176000} + \frac{379740674928681\,{\tau}^8}{802816000} - 
  \frac{189083279254923\,{\tau}^9}{28672000}
\right)+
\nonumber
\\
& n^{10} & \left( 
\frac{155277}{3276800} + \frac{911979\,\tau }{819200} + \frac{36721917\,{\tau }^2}{3276800} + \frac{110699379\,{\tau }^3}{1638400} + 
  \frac{1837467099\,{\tau }^4}{6553600} + \frac{2973858543\,{\tau }^5}{3276800} 
+ \right.
  \nonumber
  \\
  & & \left.
+\frac{76627955097\,{\tau }^6}{32768000} + 
  \frac{104357662929\,{\tau }^7}{16384000} + \frac{853398339489\,{\tau }^8}{131072000}
\right)+
\nonumber
\\
& n^{11} & \left( 
- \frac{829433601}{5046272000}   - \frac{22603330989\,\tau }{5046272000} - \frac{7219643481\,{\tau }^2}{180224000} - 
  \frac{20133254457\,{\tau }^3}{57344000} - \frac{1149209550873\,{\tau }^4}{2523136000}
+ \right.
  \nonumber
  \\
  & & \left.
 - \frac{218994700592277\,{\tau }^5}{12615680000} + 
  \frac{1073108844538299\,{\tau }^6}{6307840000} - \frac{61281594304289307\,{\tau }^7}{22077440000}
\right)+
\nonumber
\\
& n^{12} & \left( 
\frac{131410269}{183500800} + \frac{2903445891\,\tau }{183500800} + \frac{4533651\,{\tau }^2}{25600} + \frac{22510325169\,{\tau }^3}{18350080} 
+ \right.
  \nonumber
  \\
  & & \left.
+ 
  \frac{1165811276367\,{\tau }^4}{183500800} + \frac{5189828163921\,{\tau }^5}{229376000} + \frac{89579196304317\,{\tau }^6}{917504000}
\right)+
\nonumber
\\
& n^{13} & \left( 
- \frac{15299148393873}{5651824640000}  - \frac{14089860473859\,\tau }{209924915200} - \frac{330886579531671\,{\tau }^2}{565182464000} - 
  \frac{13990469422488399\,{\tau }^3}{1836843008000} 
+ \right.
  \nonumber
  \\
  & & \left.
+ \frac{20015370592779843\,{\tau }^4}{1335885824000} - \frac{9384066047520578313\,{\tau }^5}{10496245760000}
\right)+
\nonumber
\\
& n^{14} & \left( 
\frac{66042560169}{5138022400} + \frac{580058908857\,\tau }{2569011200} + \frac{15101741931291\,{\tau }^2}{5138022400}
+ \right.
  \nonumber
  \\
  & & \left.
 + 
  \frac{125170590832281\,{\tau }^3}{6422528000} + \frac{7885818877083003\,{\tau }^4}{51380224000}
\right)+
\nonumber
\\
& n^{15} & \left( 
-\frac{1557529661529486369}{29389488128000000}   - \frac{13663672258178594727\,\tau }{14694744064000000} - 
  \frac{107529809054090820291\,{\tau }^2}{14694744064000000} 
+ \right.
  \nonumber
  \\
  & & \left.
- \frac{46100805957050412573\,{\tau }^3}{262406144000000}
\right)+
\nonumber
\\
& n^{16} & \left( 
\frac{190687314873528513}{723433553920000} + \frac{320621966627776497\,\tau }{180858388480000} + \frac{18912071856450181023\,{\tau }^2}{361716776960000}
\right)+
\nonumber
\\
& n^{17} & \left( 
- \frac{126373462658844234883011}{111915170791424000000}   - \frac{92659942781039607442731\,\tau }{55957585395712000000}
\right)+
\nonumber
\\
& n^{18} &
\frac{10254234479592769713}{1808583884800000}
\label{sern}
\eeqa
\end{widetext}

\section{The Variational Free Energy and the plateau of $q(x)$}
\label{xplat}

In this appendix we consider the effect of a small plateau in the function $q(x)$ at small values of $x$.
Such a plateau in the function $q(x)$ is present in two notable cases: i) when a magnetic field is present and ii) when the parameter $n$ is finite. 
In both cases the perturbative solution in power series near the critical temperature shows that the dependence of the plateau and of the free energy on the small perturbations ({\it i.e.} the value of $n$ or $h$) is anomalous: i) we have seen in the previous sections that the behaviour of $\Phi(n)$ at $h=0$  is $\Phi(n)=f_{eq}+O(n^5)$  and ii) it is well-known that the fourth derivate of the free energy ($n=0$) with respect to the field is divergent $f(h)=f_{eq}(0)-h^2/2+O(h^{10/3})$ \cite{MPV,CRT}.
The origin of this can be traced back to the fact that the variation of the free energy in presence of a small plateau is a fifth order effect. Here we show that this can be proved at all temperatures.

We also note that the fact that the presence of a small plateau of eight $q_0$ gives an $O(q_0^5)$  correction to the free energy provides further insight into one of the earliest observations on the RSB solution \cite{POLD}, namely the fact that the corrections to the free energy due to using a finite number $K$ of RSB steps decrease like $K^{-4}$. Indeed the function $q(x)$ in this case is a set of $K$ small plateaus with $O(K^{-1})$ differences from the true solution and it is natural to expect the total free energy correction to be $O(K^{-5}) \times K$.    

We consider the variation of the free energy functional $\Phi(n,h,q(x))$ as a function of $n$, $h$ and $q_0$ under the assumption that $q(x)$ is unperturbed for $x>x_{plateau}$ (where $x_{plateau}$ is such that $q(x_{plateau})=q_0$) while $q(x)=q_0$ for $n<x<x_{plateau}$. The main result of this appendix is the following expression {\it valid at any temperature}:
\begin{displaymath}
 \Phi(n,h,q_0)-\Phi(0,0,0)=  -{h^2 \over 2}-{h^2 n \beta q_0 \over 2}-{n^2 \beta^2 q_0^3 \over 6}+
\end{displaymath}
\beq
-{\beta^2 q_0^5 \over 15 \dot{q}^2(0)} + {\beta h^2 q_0^2 \over 4 \dot{q}(0)}+ {5 \beta^2 n q_0^4 \over 24 \dot{q}(0)}+{\rm sixth \  order\  terms.} 
\label{rmain}
\eeq
Where $\dot{q}(0)$ is the derivative of the Parisi solution $q(x)$ in $x=0$ for $h=n=0$. 
The meaning of the last term is that this expression is valid at all temperature but at the lowest orders in $h$, $n$ and $q_0$.
The first term $-h^2 / 2$  yields the known result that the zero-magnetic-field susceptibility is equal to one in the whole spin-glass phase while the remaining terms are fifth order in $q_0$ in the sense that $n=O(q_0)$ and $h^2=O(q_0^3)$. 

Extremizing the above expression with respect to $q_0$ at $h=0$ we get:
\beq
q_0={3 \over 2} n \dot{q}(0)+o(n)
\eeq
and \footnote{The fact that the $O(n^6)$ term is missing in the expansion of $\Phi(n)$ (see eq. (\ref{sern})) cannot be recovered from eq. (\ref{rmain}) which is only valid at the fifth order} 
\beq
\Phi(n)=f_{eq}-{9 \over 640} (n \beta)^5 (T \dot{q}(0))^3+O(n^7)
\label{exn}
\eeq
Conversely, extremizing with respect to $q_0$ at $n=0$ we get:
\beq
q_0=\left( {3 T \dot{q}(0) \over 2 } \right)^{1/3} h^{2/3}+o(h^{2/3})
\eeq
and 
\beq
f(h)=f(0)-{h^2 \over 2}+{3 \over 20}\left( {9  \over 4 T \dot{q}(0) } \right)^{1/3} h^{10/3}+o(h^{10/3})
\label{exh}
\eeq
Thus the anomalous behaviour of the free energy at small $n$ or $h$ found near the critical temperature holds true at all orders, and the coefficients of the terms $O(n^5)$ and $O(h^{10/3})$ in the above expressions depend on the temperature only through the term $T \dot{q}(0)$, that has a finite limit at zero temperature.

The quantity $\dot{q}(0)$ can be computed in power series near the critical temperature \cite{CR} and reads:
\begin{widetext}
\beqa
\dot{q}(0) & = &\frac{1}{2} + \frac{3\,\tau}{2} + 2\,\tau^3 - 9\,\tau^4 + \frac{336\,\tau^5}{5} - 481\,\tau^6 + 
  \frac{136884\,\tau^7}{35} - \frac{979779\,\tau^8}{28} + \frac{71633011\,\tau^9}{210} - 
  \frac{1077802999\,\tau^{10}}{300} 
\nonumber
\\
& + &  \frac{18770216489\,\tau^{11}}{462} - 
  \frac{68028264769963\,\tau^{12}}{138600} + \frac{1136615361900763\,\tau^{13}}{180180} - 
  \frac{1084041597207443333\,\tau^{14}}{12612600} + 
\nonumber
\\
& + &\frac{117077323215309512399\,\tau^{15}}{94594500} - 
  \frac{4061851935671767738451\,\tau^{16}}{216216000} + 
  \frac{551046886980280618398589\,\tau^{17}}{1837836000} 
\nonumber
\\
&-& 
  \frac{1162702256772757485034973381\,\tau^{18}}{231567336000} + 
  \frac{193682918656993987102843106053\,\tau^{19}}{2199889692000}+O(\tau^{20})
\eeqa
\end{widetext}
This expression can be resummed through Pad\'e approximants \cite{CR} in order to obtain quantitative predictions in the whole low-temperature phase, {\it e.g.} in the zero temperature limit we have \cite{CR}:
\beq
\lim_{T \rightarrow 0} T \dot{q}(0)=0.743 \pm 0.002
\eeq
A more precise estimate $\lim_{T \rightarrow 0} T \dot{q}(0)=.743368$ was obtained recently working directly at $T=0$ in \cite{Opper}.
From the power series expression of $\dot{q}(0)$ the corresponding power series of the coefficients of the $O(n^5)$ and $O(h^{10/3})$ terms in the expressions (\ref{exn}) and (\ref{exh}) can be computed and they are in full agreement with the corresponding expressions computed through the power series solution of the variational equations, {\it i.e.} eq. (\ref{sern}) above and eq.(10) of Ref. \cite{CRT}.

The variational expression eq. (\ref{rmain}) can be obtained by computing the function $f(x,y)$ as a power series of $x$ around $x=0$ using the evolution equations (\ref{eqf}) up to the fifth order.
To obtain the result we need to use the fact that $f(x,y)$ is an even function of $y$ and most importantly the following three exact statements  concerning the functions $q(x)$ and $f(x,y)$ computed at $n=h=0$ (see Ref. \cite{CR} for their derivation):
\beq
\ddot{q}(0)  =  0
\eeq
\beq
f^{(0,2)}(0,0)  =  1
\eeq
\beq
f^{(0,4)}(0,0)  =  -\sqrt{ {2 \over T \dot{q}(0)} }
\eeq

\end{document}